\documentclass{article}

\PassOptionsToPackage{numbers, compress}{natbib}

\usepackage[final]{tmp}

\usepackage[utf8]{inputenc} 
\usepackage[T1]{fontenc}    
\usepackage{hyperref}       
\usepackage{url}            
\usepackage{booktabs}       
\usepackage{amsfonts}       
\usepackage{nicefrac}       
\usepackage{microtype}      
\usepackage{xcolor}         
\usepackage{graphicx}       
\usepackage{amsmath}	

\title{A New Task: Deriving Semantic Class Targets for the Physical Sciences}

\author{%
  Micah Bowles\thanks{\url{https://mb010.github.io/}} \\
  Jodrell Bank Centre for Astrophysics\\
  The University of Manchester\\
  Manchester, United Kingdom, M13 9PY \\
  \texttt{micah.bowles@postgrad.manchester.ac.uk}\\
  \And
  Hongming Tang\\
  Department of Astronomy\\
  Tsinghua University\\
  Beijing 100084, China\\
  \texttt{hongmingt@mail.tsinghua.edu.cn}\\
  \And
  Eleni Vardoulaki\\
  Th\"{u}ringer Landessternwarte\\
  Sternwarte 5, 07778 Tautenburg, Germany\\
  \texttt{elenivard@gmail.com}\\
  \And
  Emma L. Alexander\\
  Jodrell Bank Centre for Astrophysics\\
  The University of Manchester\\
  Manchester, United Kingdom, M13 9PY\\
  \texttt{emma.alexander@manchester.ac.uk}\\
  \And
  Yan Luo \\
  School of Physics and Astronomy\\
  Sun Yat-sen University\\
  2 Daxue Road, Zhuhai 519082, China\\
  \texttt{luoy355@mail2.sysu.edu.cn}\\
  \And
  Lawrence Rudnick\\
  Minnesota Institute for Astrophysics\\
  University of Minnesota\\
  116 Church St., SE, Minneapolis, MN 55455, USA\\
  \texttt{larry@umn.edu}
  \And
  Mike Walmsley \\
  Jodrell Bank Centre for Astrophysics\\
  The University of Manchester\\
  Manchester, United Kingdom, M13 9PY\\
  \texttt{michael.walmsley@manchester.ac.uk}\\
  \And
  Fiona Porter \\
  Jodrell Bank Centre for Astrophysics\\
  The University of Manchester\\
  Manchester, United Kingdom, M13 9PY\\
  \texttt{fiona.porter-2@manchester.ac.uk}\\
  \And
  Anna M.~M.~Scaife \\
  Jodrell Bank Centre for Astrophysics\\
  The University of Manchester\\
  Manchester, United Kingdom, M13 9PY\\
  \texttt{anna.scaife@manchester.ac.uk}\\
  \And
  Inigo Val Slijepcevic \\
  Jodrell Bank Centre for Astrophysics\\
  The University of Manchester\\
  Manchester, United Kingdom, M13 9PY\\
  \texttt{inigo.slijepcevic@postgrad.manchester.ac.uk}\\
  \And
  Gary Segal\\
  School of Mathematics and Physics\\
  University of Queensland\\
  St Lucia, Brisbane, QLD 4072, Australia\\
  \texttt{gp.segal@gmail.com}\\
}
\author{%
Micah Bowles$^{1}$\thanks{\href{mailto:micah.bowles@postgrad.manchester.ac.uk}{micah.bowles@postgrad.manchester.ac.uk}; \url{https://mb010.github.io/}} \quad Hongming Tang$^2$ \quad  Eleni Vardoulaki$^3$ \\
\textbf{Emma L. Alexander}$^1$ \quad \textbf{Yan Luo}$^4$ \quad \textbf{Lawrence Rudnick}$^5$ \quad \textbf{Mike Walmsley}$^1$\\
\textbf{Fiona Porter}$^1$ \quad \textbf{Anna M.~M.~Scaife}$^{1,6}$ \quad \textbf{Inigo Val Slijepcevic}$^{1}$ \quad \textbf{Gary Segal}$^{7,8}$\\ \\
$^{1}$Jodrell Bank Centre for Astrophysics, University of Manchester, Manchester, UK\\
$^{2}$Department of Astronomy, Tsinghua University, Beijing, China\\
$^{3}$Th\"{u}ringer Landessternwarte, Tautenburg, Germany\\	
$^{4}$School of Physics and Astronomy, Sun Yat-sen University, Zhuhai, China\\
$^{5}$Minnesota Institute for Astrophysics, University of Minnesota, Minneapolis, USA\\
$^{6}$The Alan Turing Institute, London, UK\\
$^{7}$School of Mathematics and Physics, University of Queensland, Brisbane, QLD, Australia\\
$^{8}$CSIRO Space and Astronomy, Epping, NSW, Australia\\
}

\begin{document}

\maketitle

\begin{abstract}
  We define \textit{deriving semantic class targets} as a novel multi-modal task. By doing so, we aim to improve classification schemes in the physical sciences which can be severely abstracted and obfuscating. 
  We address this task for upcoming radio astronomy surveys and present the derived semantic radio galaxy morphology class targets.
\end{abstract}

\section{Introduction}\label{sec:Introduction}
Language evolves and changes - sometimes quite quickly. When a new idea arises and demands new terminology, terms are commonly invented \citep[e.\,g. utopia;][]{Romm1991Utopia}, named after an early adopter or pioneer (e.\,g. Newtonian physics), or adopted from similar ideas \citep[e.\,g. `modern' in philosophy and art;][]{Schelling1994modernPhilosophy,Thomas2022art-definition}. This is also common across various physical sciences, where use of language is key, as it is even believed to affect how we think \citep[][]{Wolff2010LinguisticRelativity}.

In computer science applications, language used to describe target classes is not a pressing issue at this point. Consider ImageNet \citep[][]{deng2009imagenet}, whose class targets are built on WordNet \citep[][]{Fellbaum1998WordNet}. WordNet is explicitly constructed around the conceptual-semantic and lexical relations of the English language. Because of this, models trained on ImageNet carry inductive biases of the semantic (meaningful) terms used as class targets. 
These inductive biases do not necessarily hold true for data sets in the physical sciences, where in fewer than one hundred years new language has been developed to capture entirely novel and crucial concepts.

The physical sciences are quickly entering large data regimes where automation is essential. This includes classification. 
Supervised classification approaches usually have definitions for class targets and terms to describe to which class a given data point belongs.

To improve supervised models, it is common in both computer science and applied fields to attempt to optimise approaches and models for a fixed set of class targets. Often, computer science led state of the art methods are implemented in the hope of surpassing previous benchmarks on a specified (fixed) supervised task. However, in dynamically evolving fields, fixing such class targets in place may not be ideal or provide meaningful results. For instance, in radio astronomy, the field has developed a detailed understanding of radio galaxies and how they form, yet the same abstract classes defined through the field's understanding in the 1970s \citep[][]{FR1974MNRAS.167P..31F} still persists.
We therefore propose that rather than optimising predictions of ineffective class targets, in certain scenarios it may be more beneficial to \textit{change the target classes} with the aim of developing more robust, generalisable and feature rich models. Consequently, in this work, we propose a task to derive semantic class targets.

Sec.~\ref{sec:Task} details the proposed task and its potential consequences for the physical sciences. Sec.~\ref{sec:Method} presents the proposed method. An application of the method to radio astronomy is presented in Sec.~\ref{sec:Application} before conclusions are drawn in Sec.~\ref{sec:Conclusion}. Code and data used in this work are available at \url{https://github.com/mb010/Text2Tag}.

\section{Task}\label{sec:Task}
To improve target classes in labelled data sets, we propose a multi-modal task which can be phrased as:
\begin{quote}
    Given a set of documents describing labelled data samples, return a set of natural language terms / phrases which capture the semantic features of the labelled data set.
\end{quote}
For any task, the derived set of class labels should be able to:
\begin{enumerate}
    \item Map the science targets,
    \item Map the semantic features of the data,
    \item Use clear (non-technical) language.
\end{enumerate}

The set of targets must be able to \textit{map to the previous set of classes}, as otherwise a given scientific community will not be able to translate classifications into the historical classes that they are used to. For example, `fur length' as the class target in the supervised task of classifying cats and dogs; although useful, it does not suffice to classify a given image back into the cat/dog scheme.

Targets which map the \textit{semantic features} of the data are ideal, as populations which contain semantic feature differences may not be captured by abstract classes.
For example, a classifier could be trained to predict features of buildings (spires, column designs, materials used, etc.) rather than architectural styles (gothic, baroque, neoclassical, brutalist etc.). This would enable the model to generalise to architectural styles not included in the abstract target classes, and could even be used to highlight designs which include hybrid elements.

The benefit of \textit{clear non-technical language} is the ability it provides experts in a given field to capture, communicate, and collaborate in and around their data. If the terms map the science targets sufficiently well, they could even replace terms reducing that community's dependence on obtuse, and sometimes inconsistent, definitions of technical terminology. It could also lower barriers to entry for inter-disciplinary research, outreach, and citizen science projects.

\section{Method}\label{sec:Method}
The methods we discuss here are based on annotations and science targets. We use the term \textit{annotations} to describe short documents which each describe a feature of a single data sample using non-technical terminology. We use the term \textit{science targets} to mean the traditional abstract classifications (or engineered features) for each annotated data sample.

There are many possible approaches to address this task. Two simple approaches include manual selection of plain English terms by a panel of experts, or using a large language model (LLM) for a zero-shot approach. We expect, given appropriate experts, that manual selection via an expert panel would be acceptable to a given community. However, expecting a panel of experts to agree on a set of plain English class targets may not be realistic depending on the background of each expert and/or their ability to distil abstract concepts into simple terms. Manual selection may also lack the reproducibility and tractability that the physical sciences should demand. Using a LLM in a zero-shot approach may work; however, it is not clear how prompts should be engineered in order to extract semantic class names at scale, or if the scientific community could be convinced to accept the results of such an opaque process.

Here we propose a third method that derives class names in a bag-of-words paradigm using a pre-trained language model with embedded tokens, and a random forest (RF). 
\textit{Conceptually}, we assign each annotation the token which most closely resembles the encoding of all `similar' annotations.
A random forest is then trained to predict the abstract science targets from the derived tokens. The newly derived class targets are determined by calculating the Shapley values (estimating feature importance) of each input token and selecting the most impactful tokens as the targets. This method only results in individual tokens for each annotation. These can be manually adjusted if their original annotations clearly refer to a more specific idea which may require multiple tokens / specific grammar to capture.

The \textit{details} of our approach are as follows. We define $\mathbf{a}_j$ as the $j^{th}$ annotation of the $M$ annotations in our corpus.
Using a pre-trained model, $f_{\text{emb}}$, the annotations are embedded into a $k$-dimensional vector through $f_{\text{emb}}\left(\mathbf{a}_j\right) = \mathbf{v}_j \in \mathbb{R}^{k}$. This is currently implemented in a bag-of-words paradigm, where the order of the words does not affect the encoding. More explicitly, in the case of annotations longer than one word, $\mathbf{v}_j$ is the average vector of each embedded word within the annotations.

A similarity value is calculated for each pair $(i,j)\in[1,M]^2$ of the $M$ entries. The similarity between two vectors is calculated using the cosine similarity, $g_{\text{cs sim}}:\mathbb{R}^k\xrightarrow[]{}[-1,1]$. $M$ averaged vectors are then calculated according to a similarity threshold $\sigma$ as
$$\mathbf{v}_i^\prime =
\frac{1}{M^\prime} \sum_{j=1}^M
\begin{cases}
    \mathbf{v}_j  & \text{if\hspace{10pt}} g_{\text{sim}}(\mathbf{v}_i, \mathbf{v}_j) \geq \sigma \\
    0              & \text{else,}
\end{cases}
$$
where $M^\prime$ is the number of non-zero elements in the summation. The model originally used to embed the annotations, $f_{emb}$, is then used to produce the token closest to $\mathbf{v}^\prime_i$:  $f_{\text{emb}}^{-1}(\mathbf{v}^\prime_j) \approx s_q$,
where $s_q$ is the $q^{th}$ entry of all $Q$ unique derived s. Note that $Q\leq M$ as tags are derived from the $M$ annotations.

We define $\mathbf{t} = (t_1, ..., t_Q)$ as a vectorised encoding of the assigned derived class targets and train a model, $f_y$, to predict values of each science class $y \in Y$, where $Y$ is the set of considered science classes.
For each model, $f_y$, and tag representation, $t_q$, we calculate an importance
$$f_{\text{Importance}}(t_q, f_y) = I_{(q,y)} \in \mathbb{R},$$
where larger values of $I_{(q,y)}$ mean that $s_q$ is is more important to the classification output. We use Shapley values as a proxy for importance.

To recover the importance of the $q^{th}$ tag, we take the support weighted average across $I_{(q,y)} \forall y \in Y$:
$$ I_q = \frac{1}{Q}\sum_{y \in Y}n_y I_{(q,y)}, $$
where $n_y$ is the number of positive occurrences of class $y$. We normalise the importance values, $I_q$, such that
$$ \sum_{q=1}^{Q}I^\prime_{q} = 1, \text{\hspace{10pt}and\hspace{10pt}} I^\prime_{q} \in [0,1].$$
Finally, we sort the tags, $S_q$, according to their respective $I^\prime_q$. 
While most tags will have non-zero $I^\prime_q$ in this ranking, the majority of the information is seen to be contained within the highest ranked tags. Furthermore, the lowest ranked tags are expected to contain significant noise. We recommend making a selection cut on the minimum importance a tag can have to be included in a final $Q^\prime<Q$ tags. These most important tags, consisting of $Q^\prime$ strings $s\in S_{\text{targets}}$, are the derived semantic class targets.

Some of the tags may require clarification to improve their clarity. If a tag is not immediately clear, we suggest considering the raw annotations from which that tag was derived in order to verify what it represents and make adjustments as, or if, necessary.

\section{Application to Radio Galaxy Morphology}\label{sec:Application}
Radio galaxy classification schemes are complex, require explicit definition as their use can vary from person to person, and rely on a shared set of features. However, these classifications are essential for studying populations of radio galaxies and their properties in order to probe physical environments and conditions which cannot be recreated on earth. Working towards the Radio Galaxy Zoo EMU citizen science project (RGZ EMU; Tang \& Vardoulaki et al. in prep), the RGZ EMU team collected annotations and science classifications for 299 data samples. These samples consist of radio maps of galaxies as well as optical \citep[DSS;][]{lasker1990guide} and infra-red \citep[WISE;][]{Wright2010WISE} backgrounds to help users understand which radio components belong together. The radio maps are cutouts from the Evolutionary Map of the Universe \citep[EMU;][]{Norris+2021a,Norris+2021b} pilot survey and were subject to selection effects designed to sample galaxies with interesting morphologies from simple catalogues.

In total, 5 experts made 1,257 multi-label classifications using provided science targets. The presented science targets were selected from a compiled list of radio galaxy morphology classes presented in \citep[][]{Rudnick+2021}. These are: Single, Double, Classical double, Triple, Narrow-angle tail (NAT), Wide-angle tail (WAT), Bent tail, Fanaroff \& Riley Class 1 (FR I), Fanaroff \& Riley Class 2 (FR II), Fanaroff \& Riley Class 0 (FR 0$^*$), Hybrid, X-shaped, S-shaped, C-shaped, Diffuse, Double-double (DDRG), Core-dominant, Core-jet, Compact Symmetric Object (CSO), 1-sided, Odd Radio Circle (ORC), and Star Forming Galaxy (SFG). The plain English annotations were made by 19 users, who collectively entered 2,920 descriptions of the presented samples for a total of 8,486 annotations.

\begin{figure}
    \centering
    \includegraphics[width=\linewidth]{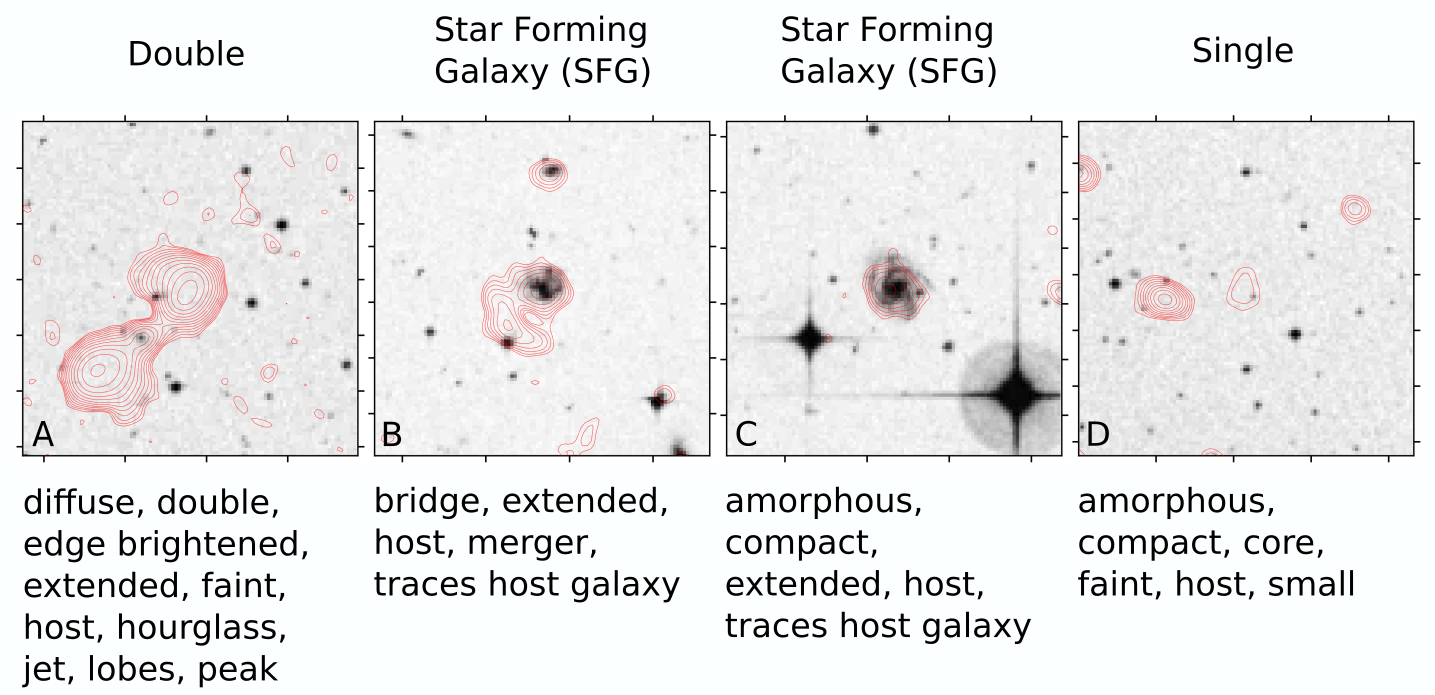}
    \caption{Example of science (above) and semantic (below) classifications in radio galaxy morphology. The classes all describe radio morphology of the source associated with the central emission in each frame. Radio EMU pilot data (red contours) overlayed on optical DSS data. We make no statements on the accuracy of these classes, but simply present them}
    \label{fig:application examples}
\end{figure}

By applying the approach presented in Sec.~\ref{sec:Method} and making hyperparameter selections based on the performance of the random forests through cross validation, we derive 22 terms. In alphabetical order, these are:
amorphous, asymmetric brightness, asymmetric structure, bent,
bridge, compact, core, diffuse, double, edge brightened, extended,
faint, host, hourglass, jet, lobe, merger, peak, plume, small, tail, and
traces host galaxy. These will be used in the RGZ EMU project and in the value added EMU catalogues. 
We present four sources in Fig.~\ref{fig:application examples}, highlighting the stark difference between the science classes and the newly assigned semantic classes.

\section{Conclusion}\label{sec:Conclusion}
We propose and define the novel task of deriving semantic class targets.
The task can be approached in many different domains, and could have far reaching effects across the physical sciences when addressed.

We present a tractable and reproducible method using annotations of, and respective science targets for, a given data set. Our method's implementation is domain-agnostic and publicly available: \url{https://github.com/mb010/Text2Tag}.

We present a first application of this method to radio galaxy morphology classification, where highly technical and disputed science classes are common. We derive 22 semantic plain English class targets. These will be implemented in the upcoming RGZ EMU citizen science project.

\section*{Impact Statement}
We hope that addressing this task will have far reaching impacts in the physical sciences. In data-driven scientific domains, the science targets may be a severe limitation in deep learning, where representations are learned to map targets, even if they are abstracted classes \citep[][]{Bowles2022_Equivariance}. Altering the class targets to be more semantic may be a significant advantage as models 
are required to fit the \textit{semantic features} which define the abstract. 
We expect this to result in more robust mappings of real data, where the representations which the model maps are clear features, even if in some regions the respective abstract classes might vary.

We hope that models trained in this way, will provide improvements to transfer learning tasks, robustness in representation learning, and domain specific foundation models across the physical sciences. Furthermore, in scenarios where language is a barrier to future research, domains within the physical sciences may benefit from partially freeing themselves from overly constrained and abstracted language to improve the accessibility and utility of complex scientific concepts for all parties including domain experts, inter-disciplinary collaborators, and the general public. Negative impacts of this work may be assumed effectiveness leading to the loss of valuable scientific information through the oversimplification of classification targets.

\begin{ack}
MB, MW, ELA, AMS, and IVS gratefully acknowledge support from the UK Alan Turing Institute under grant reference EP/V030302/1.
HT gratefully acknowledges the support from the Shuimu Tsinghua Scholar Program of Tsinghua University. EV acknowledges support by the Carl Zeiss Stiftung with the project code KODAR. ELA additionally gratefully acknowledges support from the UK Science \& Technology Facilities Council (STFC) under grant reference ST/P000649/1.

This work has made use of data from the European Space Agency (ESA) mission
{\it Gaia} (\url{https://www.cosmos.esa.int/gaia}), processed by the {\it Gaia}
Data Processing and Analysis Consortium (DPAC,
\url{https://www.cosmos.esa.int/web/gaia/dpac/consortium}). Funding for the DPAC
has been provided by national institutions, in particular the institutions
participating in the {\it Gaia} Multilateral Agreement.

This research has made use of "Aladin sky atlas" developed at CDS, Strasbourg Observatory, France.

The Australian SKA Pathfinder is part of the Australia Telescope National Facility which is managed by CSIRO. Operation of ASKAP is funded by the Australian Government with support from the National Collaborative Research Infrastructure Strategy. ASKAP uses the resources of the Pawsey Supercomputing Centre. Establishment of ASKAP, the Murchison Radio-astronomy Observatory and the Pawsey Supercomputing Centre are initiatives of the Australian Government, with support from the Government of Western Australia and the Science and Industry Endowment Fund. We acknowledge the Wajarri Yamatji as the traditional owners of the Observatory site.

We thank the citizen scientists for their time and effort.

\end{ack}

\small
\bibliographystyle{abbrv}
\bibliography{bibliography}

\end{document}